\documentclass{article}
\usepackage{kmn,latexsym,subfigure,epsfig,graphics}
\newcommand{\mc}{\multicolumn}

\newcommand{\gtsim}{\mbox{{\raisebox{-0.4ex}{$\stackrel{>}{{\scriptstyle\sim}}
$}}}}
\newcommand{\ltsim}{\mbox{{\raisebox{-0.4ex}{$\stackrel{<}{{\scriptstyle\sim}}
$}}}}

\newcommand{\figstart}[1]
    { \begin{figure}[htb]
      \begin{picture}(0,#1) }
\newcommand{\figend}[4]
    { \end{picture}
      \special{#1}
      \caption[#2]{#3}
      \label{#4}
      \end{figure} }

\def \cosone{$\Omega_{\rm M}=1$ and $\Omega_{\Lambda}=0$}
\def \costwo{$\Omega_{\rm M}=0.3$ and $\Omega_{\Lambda}=0.7$}
\def \Punit{W\,Hz$^{-1}$sr$^{-1}$}
\begin{document}

\title[The steep-spectrum redshift cut-off]{On
the redshift cut-off for steep-spectrum radio sources}

\author[Jarvis at al.]{Matt J.\,Jarvis$^{1,2}$\thanks{Email: jarvis@strw.leidenuniv.nl}, 
Steve Rawlings$^{1}$, Chris J.\ Willott$^{1}$,\\
\vspace{-1.5mm}\\
{\LARGE\rm  Katherine M.\,Blundell$^{1}$, Steve Eales$^{3}$ and
Mark Lacy$^{1,4,5}$}\\
\vspace{-1.5mm}\\
$^{1}$Astrophysics, Department of Physics, Keble Road, Oxford, OX1
3RH, UK \\
$^{2}$Sterrewacht Leiden, Postbus 9513, 2300 RA Leiden, the
Netherlands \\
$^{3}$Department of Physics and Astronomy, University of Wales 
College of Cardiff, P.O. Box 913, Cardiff, CF2 3YB, UK\\
$^{4}$Institute of Geophysics and Planetary Physics, L-413 Lawrence
Livermore National Laboratory, Livermore, CA 94550, USA \\
$^{5}$Department of Physics, University of California, 1 Shields
Avenue, Davis CA 95616, USA}
\maketitle

\begin{abstract}
We use three samples (3CRR, 6CE and 6C*) selected at low radio
frequency to constrain the cosmic evolution in the radio luminosity
function (RLF) for the `most luminous' steep-spectrum radio
sources. Although intrinsically rare, such sources give the largest
possible baseline in redshift for the complete flux-density-limited
samples currently available.  Using parametric models to describe the
RLF which incorporate distributions in radio spectral shape and linear
size as well as the usual luminosity and redshift, we find that the
data are consistent with a constant comoving space density between $z
\sim 2.5$ and $z \sim 4.5$. We find this model is favoured over a
model with similar evolutionary behaviour to that of
optically-selected quasars (i.e.  a roughly Gaussian distribution in
redshift) with a probability ratio of $\sim 25$\,:\,1 and $\sim
100$\,:\,1 for spatially-flat cosmologies with $\Omega_{\Lambda} = 0$
and $\Omega_{\Lambda} = 0.7$ respectively. Within the uncertainties,
this evolutionary behaviour may be reconciled with the shallow decline
preferred for the comoving space density of flat-spectrum sources by
Dunlop \& Peacock (1990) and Jarvis \& Rawlings (2000), in line with
the expectations of Unified Schemes.

\end{abstract}

\begin{keywords}
galaxies:active - galaxies:luminosity function, mass function
- radio continuum:galaxies
\end{keywords}

\section{Introduction}\label{sec:intro}
Speculation concerning the cosmic evolution of the radio source
population has been at the forefront of astrophysical research since
the 1960s. It soon became clear that the
radio luminosity function (RLF) declines steeply towards high radio
luminosities, and that the (comoving) space densities of the rarest,
most luminous quasars and radio galaxies were much higher at epochs
corresponding to $z \sim 2$ than they are now (Longair 1966).  As
emphasised by Peacock (1985), Dunlop \& Peacock (1990; hereafter
DP90), Shaver et al. (1996; hereafter SH96), Jarvis \& Rawlings (2000;
hereafter JR00) and many others the crucial advantage of any radio-based work is
that with sufficient optical follow-up, it can be made free of optical
selection effects, such as increasing dust obscuration at high
redshift.

The term `redshift cut-off' made an early appearance in the literature
(e.g. Sandage 1972), and is now taken to mean any significant decline
in the space density over $2.5\, \ltsim\,z\,\ltsim\, 5$. Peacock (1985) and then DP90, found evidence for a redshift
cut-off in the distribution of flat-spectrum radio quasars over the
redshift range $2-4$.  Through failing to find any flat-spectrum radio
quasars at $z >5$ in a survey covering $\sim\!4$\,sr, SH96 (see also
Shaver et al. 1998; hereafter SH98) argued for a drop in space density
between $2.5\, \ltsim\,z\, \ltsim\,6$ of more than 1\,dex. JR00
re-addressed this question using a similar sample to the flat-spectrum
sample of SH96/SH98. By considering the effects of a distribution in
spectral index and also the presence of any curvature in the radio
spectra, JR00 found evidence for a decline between $2.5 \leq z \leq 5$
by a factor $\sim 4$ in agreement with DP90, although they found that
steeper declines such as that preferred by SH96, as well as models
invoking a constant comoving space density at high redshift, can only
be ruled out at the 90\% confidence level with the surveys currently
available.

DP90 also claimed the first evidence for a similar redshift cut-off in
the steep-spectrum radio population. According to simple Unified
Schemes (e.g. Barthel 1989; Antonucci 1993) steep-spectrum radio sources are the
parent, unbeamed, population for Doppler-boosted flat-spectrum
objects, and should show similar evolutionary behaviour (Jackson \&
Wall, 1999)\footnote{JR00 have also suggested that Giga-Hertz Peaked
Spectrum (GPS) sources also play a major role in the flat-spectrum
surveys along with Doppler Boosted sources, particularly at the high luminosity end of the RLF.}. However, the high selection frequency
(2.7\,GHz) of the steep-spectrum samples used by DP90 meant that their
survey was sensitive to only the most luminous sources at
high redshift, an effect which is made more pronounced by the positive
correlations between spectral index and redshift/radio luminosity
(e.g. Blundell, Rawlings \& Willott 1999). The necessity of considering the effects of the
distribution in radio spectral parameters, highlighted by JR00 for the
flat-spectrum population, is especially acute for this population which
typically have steep and curved spectra. There are also worries about
the incompleteness of the faint radio-selected sample used by DP90
(see their Appendix A).  Moreover, DP90 were forced to use photometric
redshift estimates for most of their high-redshift steep-spectrum
objects, and since these were based on the still poorly-understood
$K-z$ diagram for radio galaxies (e.g. Eales et al. 1997; Jarvis et
al. 2001b), this leads
to considerable uncertainty in their results.

To ameliorate these problems, our group has been seeking spectroscopic
redshifts for samples selected at low radio frequency from the 6C and 7C
catalogues (see e.g. Rawlings et al. 1998). There are now three
low-frequency-selected samples with essentially complete redshift
information: 3CRR (Laing, Riley \& Longair, 1983); 6CE (Eales et
al. 1985; Rawlings, Eales \& Lacy, 2001) and the 7C redshift survey
(7CRS; Blundell et al. in prep.; Willott et al. in prep.).
Investigation of the RLF
of steep-spectrum quasars (Willott et al. 1998), the first of its
kind, found no definitive evidence for a redshift cut-off, albeit with
only a small sample of quasars. 
Using the combined 3CRR, 6CE and 7CRS dataset, comprising 357 sources,
Willott et al. (2001) obtained similar results for the entire
low-frequency population.

In this paper we focus on whether there is any evidence for a redshift
cut-off in the space density of the most radio-luminous
steep-spectrum radio sources. Our work on follow-up spectroscopy of
the 6CE (Rawlings et al. 2001) and 6C*
(Blundell et al. 1998; Jarvis et al. 2001a, 2001b) samples provides
redshifts for well defined samples containing the distant counterparts
of the most luminous radio sources in the 3CRR survey (Laing, Riley \&
Longair 1983). This analysis of the RLF is different from that by
Willott et al. (2001) since it also includes 6C*, which is crucial in
sampling to high redshift.

In Sec.~\ref{sec:complete_samples} we summarise the relevant
properties of the 3CRR, 6CE and 6C* samples, and describe how the most
luminous radio sources have been drawn from them. In
Sec.~\ref{sec:most_luminous} we highlight the reasons for
concentrating on just the most radio-luminous sources and in
Sec.~\ref{sec:modellingRLF} we describe the parametric models used in
a maximum likelihood analysis which yield estimates of the RLF. We
use the banded $V / V_{\rm max}$ statistic in Sec.~\ref{sec:vvmax} to
compare our parametric modelling with a non-parametric approach and in
Sec.~\ref{sec:zdist} we compare our predicted redshift distributions
with the data and highlight the importance of small number
statistics. In Sec.~\ref{sec:discussion} we discuss our results in the
context of unified schemes of radio AGN.

We assume $H_{\circ} = 50~{\rm km~s^{-1}~Mpc^{-1}}$, and obtain
all results in two spatially-flat cosmologies (\cosone, hereafter
cosmology I and
\costwo, hereafter cosmology II).  We use the convention for spectral index $\alpha$ that
$S_{\nu} \propto \nu^{-\alpha}$ and $\alpha_{\nu_{1}}^{\nu_{2}}$
represents the spectral index between observed frequencies $\nu_{1}$
and $\nu_{2}$. All luminosities quoted are measured in units of \Punit
at rest-frame 151\,MHz unless stated otherwise.

\begin{figure} 
{\hbox to 0.45\textwidth{ \null\null \epsfxsize=0.45\textwidth
\epsfbox{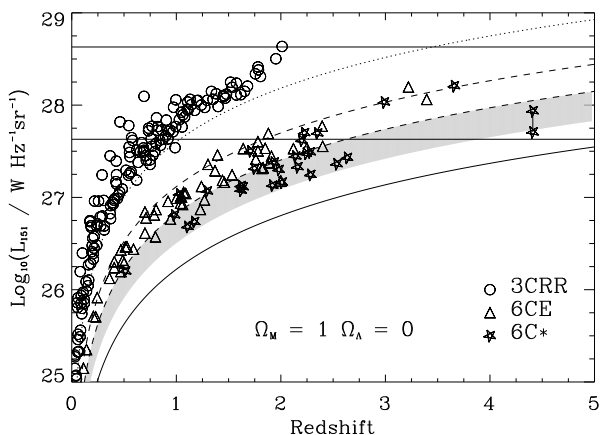} }} 
%\vspace{1cm} {\hbox to \textwidth{ \null\null
%\epsfxsize=0.75\textwidth \epsfbox{\where/pzcos4.ps} }}
{\caption{\label{fig:pzplane} Rest-frame 151\,MHz luminosity
($L_{151}$) versus redshift $z$ plane for the three samples used in
our analysis in cosmology I; 3CRR (circles),
6CE (triangles) and 6C* (stars). The rest-frame 151\,MHz luminosity
$L_{151}$ has been calculated according to the polynomial fit to the
radio spectrum described in Sec.~\ref{sec:specshape}.  The 6C* source
6C*0140+326 at $z=4.41$ is represented by two points joined by a solid
vertical line. This is because 6C*0140+326 is thought to be subject to
gravitational lensing by a foreground galaxy (Rawlings et al. 1996)
and the highest luminosity point corresponds to the lensed luminosity
whereas the lower point corresponds to the de-magnified, intrinsic
luminosity of the source in the absence of significant lensing. The
area between the horizontal lines is the region which contains the
`most luminous' sources according to our definition. The curved lines
show the lower flux-density limit for the 3CRR (dotted) and 7CRS
(solid). The dashed lines correspond to the limits for the 6CE sample
and the shaded region shows the 6C* flux-density limits (all assuming
a radio spectral index of 0.5).  Note the area between the 3CRR
sources and 6CE sources contains no sources, this is the area which
corresponds to the absence of a flux-density-limited sample between
the 6CE ($S_{151} \leq 3.93\:$Jy) and 3CRR ($S_{178} \geq 10.9\:$Jy)
samples. }} \null
\end{figure}
 
\section{Complete samples of radio sources selected at low frequency}\label{sec:complete_samples}
\subsection{The advantages of using three samples}\label{sec:3samples}
\begin{table*}
\begin{center}
\begin{tabular}{llccccl}
\hline\hline
\mc{1}{l|}{Sample} & \mc{1}{c|}{Flux-density} & \mc{1}{c|}{Sky Area} &
\mc{1}{c|}{Number of} & \mc{1}{c|}{Uncertain} & \mc{1}{c|}{Spectroscopic} &\mc{1}{c|}{Reference} \\
\mc{1}{l|}{} & \mc{1}{c|}{limits / Jy} & \mc{1}{c|}{/ sr} &
\mc{1}{c|}{Sources} & \mc{1}{c|}{redshifts} & \mc{1}{c|}{completeness}
& \mc{1}{c|}{} \\
\hline\hline
3CRR & $S_{178}> 10.9$ & 4.24 & 170 & 0 & 100\% & Laing et al. (1983)  \\
6CE  & $2.00 \leq S_{151} \leq 3.93$ & 0.10 & 58 & 7 & 97\% & Rawlings et al. (2001) \\
6C* & $0.96 \leq S_{151} \leq 2.00$ & 0.13 & 29 & 6 & 100\% & Jarvis et al. (2001a) \\
\hline\hline
\end{tabular}
{\caption{\label{tab:samples} Summary of the three samples used in the
analysis to constrain the RLF of the most luminous radio sources. The
spectroscopic completeness figure neglects uncertainties in a small
number of redshifts in the 6CE and 6C* samples. }}
\end{center}
\end{table*}
Flux-density-limited samples over a particular sky area provide a
direct way of constraining the RLF if they have complete redshift
information. 
%The principal uncertainties arising in previous
%determinations of the RLF using low-frequency-selected samples are
%usually due to the lack of a spectroscopically-complete sample below
%the flux-density limit of the 3CRR survey. With the recently completed
%samples 6CE, 7CRS and 6C* at fainter flux-densities we are now able to
%probe a much greater area of the luminosity - redshift plane and
%decouple the correlations between luminosity and redshift inherent in
%any single flux-density-limited sample. 
In this analysis, in which
only the `most luminous' sources are considered, the 7CRS with a
flux-density limit $S_{151} \geq 0.5$\,Jy covering a total sky area of
0.022\,sr would contribute only one such source, principally due to
the small sky area it surveyed and the steepness of the RLF (see
Willott et al. 2001). Thus, inclusion of 7CRS was deemed unnecessary
for the purpose of this paper and we only use three samples, 3CRR, 6CE
and 6C* which we describe in the next three sections.

\subsection{3CRR}\label{sec:3C}
We use the 3CRR sample of Laing, Riley \& Longair (1983), which
contains 173 sources. Following Blundell et al. (1999) we exclude
3C\,231 (M82) since the radio emission is a consequence of
star-formation (Condon et al. 1990) rather than an active nucleus. Two
other sources are also excluded (3C\,345 and 3C\,454.3) because their
emission only exceeds the 178\,MHz flux-density limit because of
Doppler boosting of their core emission. The sample has 100\%
spectroscopic redshifts.
 
\subsection{6CE}\label{sec:6CE}
The 6CE sample originally defined by Eales et al. (1985), and recently
revised by Rawlings et al. (2001), probes flux-densities a factor
$\sim 6$ fainter than 3CRR. The sample consists of 59 objects,
however, one of these (6CE1036+3616) is excluded for the statistically
unbiased reason of there being a bright star occluding any likely
optical identification. The sample considered here
consists of 58 objects, with only two objects lacking a spectroscopic
redshift, a completeness of $\sim 97$\%, with a further 7 redshifts not
yet secure.

\subsection{6C*}\label{sec:6C*}
The 6C* sample (Blundell et al. 1998) is a filtered sample from part
VI of the 6C survey of radio sources at 151\,MHz which was designed to
find radio sources at $z > 4$. The completeness of the sample is
summarised in Table~\ref{tab:samples}. At the faint flux-density limit
of 6C*, the source counts of radio sources at 151\,MHz are of order
$10^{4}$ per steradian, making it very difficult to construct a
complete sample with spectroscopically determined redshifts over large
solid angles. Moreover, the low-luminosity and thus low-redshift
sources supply a significant contribution to the source counts at
these flux-density limits making it increasingly difficult to use
faint flux-density-limited samples to probe evolution of radio sources
at high redshift. To combat this problem, additional selection
criteria based on radio properties can be used to filter out the
low-luminosity (low-redshift) sources. These additional selection
criteria are discussed in detail in Blundell et al. (1998), but to
summarise, only sources with spectral indices steeper than
$\alpha_{0.151{\rm G}}^{4.85{\rm G}} \geq 0.981$, and angular size
$\theta < 15$\,arcsec are included. These selection criteria introduce
additional complications in any analysis where the 6C* sample is used,
and need to be treated with careful consideration as will be discussed
in Sec.~\ref{sec:6C*missing}. This sample has now been reduced from 34
to 29 objects since images from the NVSS survey (Condon et al. 1998)
have confirmed the suggestions of Blundell et al. (1998) that these
objects were the hotspots of large ($\theta > 15^{\prime\prime}$)
double sources. The optical spectroscopic follow-up of this sample has
now been completed (Jarvis et al. 2001a) and of the remaining objects,
23 have secure spectroscopic redshifts, and 6 have tentative
redshifts. The redshifts of these six sources are typically based on a
single definite emission line identified with a specific feature on
the basis of a redshift estimate from the $K-$band magnitude.

It is also worth noting that the absence of strong Ly$\alpha$ emission
in the objects from 6CE and 6C* which lack secure redshifts implies
that these source are probably below $z \leq 1.8$, and should not
affect the analysis in this paper (as we are only considering the most
luminous sources which correspond to high redshifts at these
flux-densities).

\begin{table*}
\begin{center}
\scriptsize
\begin{tabular}{l|c|c|c|c|r|r|r|c}
\hline\hline
\mc{1}{l|}{Source} & \mc{1}{c|}{$z$} & \mc{2}{c|}{$\log_{10}(L_{151}$
/ \Punit)} & \mc{1}{c|}{$a_{1}$} &
\mc{1}{c|}{$a_{2}$} & \mc{2}{c|}{$D$ / kpc} & \mc{1}{c|}{Optical} \\
\mc{1}{l|}{} & \mc{1}{c|}{} & \mc{1}{c|}{Cos I} & \mc{1}{c|}{Cos II} & \mc{1}{c|}{} & \mc{1}{l|}{}  &
\mc{1}{c|}{Cos I} & \mc{1}{r|}{Cos II} & \mc{1}{c|}{type} \\
\hline\hline
3C9          & 2.012 & 28.63 & 28.94 &  -0.47 &  -0.10 &  114.5 & 164.0 & Q \\
3C13         & 1.351 & 28.06 & 28.33 &  -0.33 &  -0.11 &  241.7 & 330.8 & G \\
3C14         & 1.469 & 28.07 & 28.35 &  -0.44 &  -0.08 &  205.5 & 284.1 & Q \\
3C22         & 0.938 & 27.73 & 27.96 &  -0.58 &  -0.05 &  206.2 & 269.2 & B \\
3C43         & 1.457 & 28.09 & 28.37 &  -0.75 &   0.00 &   21.4 & 29.6  & Q \\
3C55         & 0.734 & 27.75 & 27.95 &  -0.09 &  -0.16 &  572.9 & 723.8 & G \\
3C65         & 1.176 & 27.94 & 28.19 &   0.16 &  -0.18 &  149.7 & 201.4 & S \\
3C68.1       & 1.238 & 28.00 & 28.26 &  -0.36 &  -0.10 &  447.9 & 606.4 & Q \\
3C68.2       & 1.575 & 28.23 & 28.52 &  -0.39 &  -0.16 &  189.7 & 264.5 & G \\
3C147        & 0.545 & 27.77 & 27.94 &   0.76 &  -0.22 &   22.1 &  26.8 & Q \\
3C175        & 0.768 & 27.77 & 27.97 &  -0.92 &  -0.01 &  391.3 & 497.4 & Q \\
3C181        & 1.382 & 28.12 & 28.39 &  -0.23 &  -0.11 &   49.0 & 67.2 & Q \\
3C184        & 0.994 & 27.76 & 28.00 &   0.03 &  -0.16 &   40.8 & 53.7 & G \\
3C186        & 1.063 & 27.89 & 28.14 &   0.18 &  -0.22 &   13.7 & 18.2 & Q \\
3C190        & 1.197 & 28.04 & 28.30 &  -0.63 &  -0.04 &   57.7 & 77.8 &  Q \\
3C191        & 1.952 & 28.50 & 28.81 &  -0.78 &  -0.03 &   40.3 & 57.6 & Q \\
3C196        & 0.871 & 28.35 & 28.58 &  -0.07 &  -0.13 &   83.6 & 108.0 & Q \\
3C204        & 1.112 & 27.94 & 28.19 &  -1.18 &   0.01 &  314.2 & 419.6 & Q \\
3C205        & 1.534 & 28.19 & 28.48 &  -0.34 &  -0.11 &  153.5 & 213.4 & Q \\
3C208        & 1.109 & 28.03 & 28.28 &  -0.26 &  -0.13 &   94.4 & 126.1 & Q \\
3C212        & 1.049 & 27.92 & 28.16 &  -0.78 &   0.00 &   77.0 & 102.0 & Q \\
3C217        & 0.897 & 27.66 & ----- &  -0.33 &  -0.11 &  100.8 & ----- & G \\
3C226        & 0.817 & 27.72 & 27.94 &  -0.52 &  -0.08 &  289.1 & 370.6 & G \\
3C239        & 1.781 & 28.41 & 28.71 &  -0.48 &  -0.10 &   93.7 & 132.5 & G \\
3C241        & 1.617 & 28.11 & 28.40 &   0.36 &  -0.24 &    7.7 & 10.8 & G \\
3C245        & 1.029 & 27.84 & 28.08 &  -0.67 &   0.00 &   77.7 & 102.7 & Q \\
3C252        & 1.105 & 27.95 & 28.20 &  -1.03 &  -0.01 &  514.9 & 687.1 & G \\
3C254        & 0.734 & 27.72 & 27.92 &  -0.20 &  -0.14 &  105.7 & 133.5 & Q \\
3C263.1      & 0.824 & 27.76 & 27.98 &  -0.21 &  -0.12 &   56.2 & 72.2 & G \\
3C265        & 0.810 & 27.86 & 28.07 &  -0.82 &  -0.03 &  643.3 & 823.6 & S \\
3C266        & 1.272 & 27.97 & 28.24 &  -0.34 &  -0.11 &   38.8 & 52.7 & G \\
3C267        & 1.144 & 28.00 & 28.25 &  -0.60 &  -0.05 &  326.7 & 437.9 & G \\
3C268.1      & 0.973 & 27.97 & 28.21 &  -0.70 &   0.00 &  390.5 & 512.4 & G \\
3C268.4      & 1.400 & 27.98 & 28.26 &  -0.30 &  -0.10 &   93.6 & 128.7 & Q \\
3C270.1      & 1.519 & 28.24 & 28.53 &  -0.96 &   0.02 &  102.5 & 142.2 & Q \\
3C280        & 0.996 & 28.05 & 28.29 &  -0.50 &  -0.05 &  123.4 & 162.4 & G \\
3C287        & 1.055 & 27.76 & 28.00 &   0.27 &  -0.13 &    0.8 & 1.0 & Q \\
3C289        & 0.967 & 27.71 & 27.94 &  -0.14 &  -0.12 &   84.8 & 111.2 & G \\
3C294        & 1.786 & 28.38 & 28.68 &  -0.82 &  -0.05 &  125.5 & 177.4 & G \\
3C295        & 0.461 & 27.81 & 27.96 &   0.79 &  -0.26 &   37.7 & 44.9 & G \\
3C309.1      & 0.904 & 27.83 & 28.06 &  -0.00 &  -0.10 &   24.4 & 31.7 & Q \\
3C318        & 1.574 & 28.11 & 28.40 &  -0.01 &  -0.14 & 6.8 & 9.5 & Q \\
3C322        & 1.681 & 28.20 & 28.49 &  -0.54 &  -0.07 &  278.6 & 391.2 & G \\
3C324        & 1.206 & 28.05 & 28.31 &  -0.15 &  -0.14 &   86.1 & 116.2 & S \\
3C325        & 0.860 & 27.72 & 27.95 &  -0.26 &  -0.10 &  133.4 & 172.2 & G \\
3C336        & 0.927 & 27.71 & 27.94 &  -0.68 &  -0.03 &  183.1 & 238.7 & Q \\
3C356        & 1.079 & 27.87 & 28.11 &  -0.50 &  -0.09 &  642.7 & 854.9 & G \\
4C16.49      & 1.296 & 28.05 & 28.31 &  -1.00 &   0.00 &  137.8 & 187.6 & Q \\ 
4C13.66      & 1.450 & 28.10 & 28.38 &  -0.81 &   0.00 &   51.4 & 71.0 & G \\
3C368        & 1.132 & 28.06 & 28.31 &  -0.50 &  -0.13 &   67.9 & 90.9 & G \\
3C380        & 0.691 & 28.09 & 28.29 &  -0.45 &  -0.04 &  158.8 & 199.0 & Q \\
3C427.1      & 0.572 & 27.64 & ----- &  -0.57 &  -0.07 &  209.6 & ----- & G \\
3C432        & 1.805 & 28.31 & 28.61 &  -0.38 &  -0.11 &  108.6 & 153.7 & Q \\
3C437        & 1.480 & 28.12 & 28.40 &  -0.01 &  -0.13 &  294.3 & 407.4 & G \\
3C454        & 1.757 & 28.33 & 28.63 &  -0.90 &   0.00 &   10.9 &  15.4 & Q \\
3C469.1      & 1.336 & 28.14 & 28.42 &  -1.12 &   0.00 &  636.7 & 870.3 & G \\
3C470.0      & 1.653 & 28.14 & 28.44 &  -0.39 &  -0.09 &  203.0 & 284.6 & G \\
\hline
6CE0902+3419 & 3.395 & 28.06 & 28.41 &  -0.80 &   0.00 &   27.7 & 41.4 & G \\ 
6CE0905+3955 & 1.880 & 27.69 & 27.99 &  -0.51 &  -0.10 &  920.4 & 1308.4 & G \\
6CE0929+3855 & 2.394 & 27.77 & 28.09 &  -0.60 &  -0.05 &   28.9 & 42.2 & G \\
6CE1204+3708 & 1.780 & ----- & 27.90 &   0.39 &  -0.22 &   ------ & 609.2 & G \\
6CE1232+3942 & 3.220 & 28.19 & 28 54 &   0.00 &  -0.21 &   54.4 & 81.2 & G \\
\hline
6C*0020+440  & 2.988 & 28.03 & 28.37 &  -0.78 &  -0.05 &   72.9 & 108.1 & G \\
6C*0032+412  & 3.658 & 28.20 & 28.56 &  -1.41 &   0.04 &   15.4 &  23.2 & G \\
6C*0118+486  & 2.350 & 27.69 & 28.02 &  -2.01 &   0.16 &  126.7 & 184.2 & G \\
6C*0135+313  & 2.199 & ----- & 27.93 &  -1.17 &   0.00 &   ------ &  20.8 & G \\
6C*0140+326  & 4.410 & 27.93 & 28.30 &  -0.07 &  -0.18 &   15.3 & 23.3 & G \\
6C*0142+427  & 2.225 & 27.70 & 28.02 &  -1.38 &   0.05 &  121.4 & 175.5 & G \\
\hline\hline
\end{tabular}
\end{center}
{\caption[junk]{\label{tab:sources} Sources selected from the 3CRR, 6CE
and 6C* samples which correspond to our definition of the `most radio
luminous'. The rest-frame 151\,MHz luminosities were calculated by
fitting a 1st or 2nd order polynomial to the available radio data as
described in Sec.~\ref{sec:specshape}. The classes G, Q, B and S
represent quasars, radio galaxies, broad-line radio galaxies and
sources with scattered broad lines
respectively, where we have followed the classifications of Willott et
al. (1998) and Cos I and Cos II denote cosmologies I and II
respectively. The solid horizontal lines are used where the source was not
used for the specific cosmology due to it falling below our lower
luminosity limit. }}
\end{table*}

\section{Selecting `the most luminous radio sources'}\label{sec:most_luminous}
The principal aim of this paper is to investigate the evolution in the
comoving space density of steep-spectrum radio sources directly. Thus,
we concentrate on only the `most luminous' sources, which we are able
to detect to $z\, \gtsim\, 4$ in our flux-density-limited
samples. Another reason for concentrating on just the most luminous
objects is that in such a narrow luminosity range any bias from
intrinsic correlations which are present in radio samples, such as
luminosity - spectral index and linear size - spectral index
correlations (e.g. Blundell et al. 1999) will be minimised.  Selecting
which sources from the three samples should be included in our
analysis as the `most luminous' was simply a question of where to
place the lower limit in radio luminosity. To enable us to compare
results between various cosmologies we are forced to alter the limits
between cosmologies. Therefore to make the analyses consistent we only
consider the top decade in $L_{151}$ in cosmology I (\cosone).  Thus
all sources with $\log_{10}L_{151} \geq 27.63$ were included. For
cosmology II (\costwo) we redefine our lower limit in luminosity at
$\log_{10}L_{151} \geq 27.90$, this is not strictly the lower limit of
the top-decade (which is $\log_{10}L_{151} \geq 27.95$) but allows the
same number of sources into the modelling process as in cosmology I.
%These criteria enabled us to include enough sources to minimise the
%effects of small number statistics while still excluding
%low-luminosity sources.

This selection process led to the inclusion of 57 3CRR sources for
cosmology I and 55 3CRR sources for cosmology II, four (five) 6CE
sources, and five (six) 6C* sources in cosmology I (cosmology II). 
These sources are tabulated in Table~\ref{tab:sources}. 

\section{Modelling the RLF}\label{sec:modellingRLF}
\subsection{6C* - what are we missing?}\label{sec:6C*missing}
One of the principal problems in dealing with a filtered sample such
as 6C* is determining which populations are filtered out. One of the
aims of the 6C* sample was to find objects at $z > 4$. However, to use
this sample in any determination of how the comoving space density of
radio sources evolves, we need to account for the sources excluded at
all redshifts. This is crucial because if a significant number of
sources between $2 \leq z \leq 3$ are filtered out of the sample, and
objects at $z > 3$ are not, then on first inspection the results would
be very misleading as the derived comoving space density at
high redshift would be higher relative to the comoving space density
at $z \sim 2$. 
To incorporate the
effects of the selection criteria into the parametric modelling we
adopt functional forms which parameterise the distributions in spectral
shape and linear size, in addition to luminosity and redshift. These
functional forms are described in the following sections.

\subsection{Radio spectral shape}\label{sec:specshape}
Many radio spectra are known to exhibit curvature (e.g. Laing \&
Peacock 1980; Blundell et al. 1999) thus rendering the use of a
power-law spectral index inadequate in describing the radio spectra
over a large frequency baseline. As a consequence of this, determining
the rest-frame luminosity becomes less straightforward.  Thus, to
account for spectral curvature we fit the radio spectra with a
polynomial of the form
\begin{equation}\label{eqn:polyfit}
y = \log_{10} S_{\nu} = \sum_{i=0}^{N} a_{i} x^{i},
\end{equation}
where $x = \log_{10} (\nu\,/\,{\rm MHz})$. We fitted the flux-density
data using a Bayesian polynomial regression analysis which assessed
the posterior probability density function (pdf) for the required
order of polynomial fit (e.g. Sivia 1996). This is the same method as
used in Blundell et al. (1998) and Blundell et al. (1999). 
The results of this
analysis are presented in Table~\ref{tab:sources}; note that in no
cases was $N > 2$ preferred, so the fits are either 1st or 2nd order
polynomials. 

\subsection{Parametric modelling of the RLF for the most luminous
radio galaxies and quasars}\label{sec:paramodels} To extract as much
information as possible from our dataset, we use three parametric
models, with as few free parameters as possible, to estimate the RLF of
the most luminous low-frequency-selected radio sources.

Following JR00 we construct a luminosity function separable in
luminosity and redshift along with additional parameterisations
describing the distributions in spectral shape and projected linear
size.

We use a single power-law to parameterise the luminosity function
$\rho_{L} (L_{151})$, i.e.
\begin{equation}\label{eqn:rhobeta}
\rho_{L} (L_{151}) = \left (\frac{L_{151}}{L_{\circ}}\right)^{-\beta} ,
\end{equation}
where $\beta$ is a dimensionless free parameter and $L_{\circ}$ is a
normalising luminosity fixed at the lower limit of the top-decade in
luminosity, which is well above the `break' luminosity in the RLF (Willott et
al. 2001) and $L_{151}$ is the rest-frame 151\,MHz luminosity.

\begin{figure}[!ht] 
{\hbox to 0.45\textwidth{ \null\null \epsfxsize=0.45\textwidth
\epsfbox{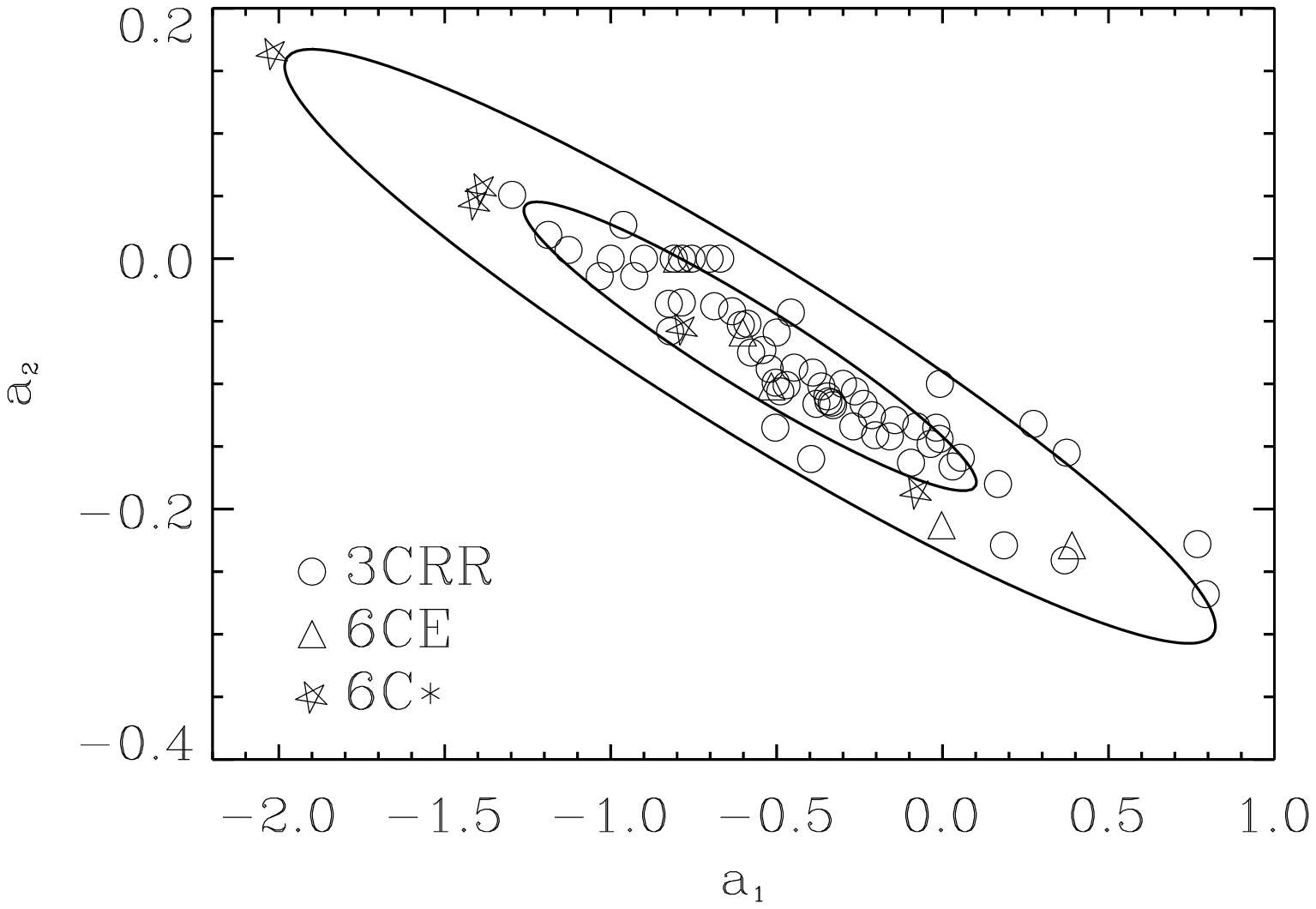} }}
{\caption{\label{fig:a1vsa2} The correlated parameters $a_{1}$ and
$a_{2}$ which describe the shape of the radio spectra. The circles are 3CRR sources; triangles correspond to 6CE sources and stars
correspond to sources from the filtered 6C* sample. The inner (outer) contour corresponds to the 1$\sigma$ (2$\sigma$)
spread about the correlation using the best-fit value of model B in
cosmology I. }}
\end{figure}

\begin{figure}[!ht] 
{\hbox to 0.45\textwidth{ \null\null \epsfxsize=0.45\textwidth
\epsfbox{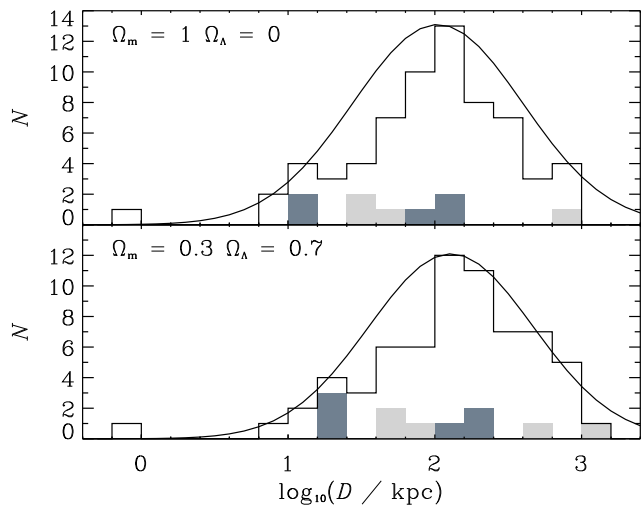}}}
%\epsfxsize=0.45\textwidth
%\epsfbox{Dhistcos2.ps} }}
{\caption{\label{fig:Dhist} Histogram of the distribution in
$\log_{10}(D / {\rm kpc})$, where $D$ is the projected linear size,
for the top-decade in luminosity for cosmology I (top) and cosmology
II (bottom). The white regions correspond to 3CRR sources, the light
shaded regions are the contributions from the 6CE sample and the dark
shaded regions are the 6C* sources. The solid line represents the
Gaussian fit to the data from model A (Sec.~\ref{sec:resultspara})
with an assumed normalisation factor which is incorporated within
$\rho_{\circ}$ in the parametric modelling. }}
\end{figure}

JR00 highlighted the importance of a distribution in spectral index in
their parametric modelling and also suggested that curvature would
need to be incorporated in a full analysis of the RLF. The small
number of sources in the JR00 sample restricted the models to a small
number of free parameters and the effects of spectral curvature could
not be incorporated into their parametric models.  However, we are
able to introduce a distribution for spectral curvature into the
modelling of the RLF here, due to the larger number of sources in the
top-decade of luminosity from our set of low-frequency-selected
samples.
 
We fit the radio spectra with a polynomial of the form described in
Sec.~\ref{sec:specshape}. Fig.~\ref{fig:a1vsa2} shows that the
parameters describing spectral shape, $a_{1}$ and $a_{2}$ are linearly
anti-correlated. Therefore to incorporate the effects of spectral
curvature into the modelling of the RLF we use a functional form which
accounts for this anti-correlation and the spread about it. To
parameterise $a_{1}$ and $a_{2}$ we rotate the $a_{1} - a_{2}$ plane
by the free parameter $\omega$ to transform the correlation so it
becomes parallel to the $a_{1}$ axis, i.e.
\begin{eqnarray}\label{eqn:rhoa1a2}
a_{1}^{\prime} - \mu_{a_{1}^{\prime}} = (a_{1} - \mu_{a_{1}^{\prime}})\cos(\omega) + (a_{2}- \mu_{a_{2}^{\prime}}) \sin(\omega), \nonumber \\
a_{2}^{\prime} - \mu_{a_{2}^{\prime}} = -(a_{1}-\mu_{a_{1}^{\prime}})\sin(\omega) + (a_{2}- \mu_{a_{2}^{\prime}})\cos(\omega),
\end{eqnarray}
where $a_{1}$ and $a_{2}$ are defined in Sec.~\ref{sec:specshape},
$a_{1}^{\prime}$ and $a_{2}^{\prime}$ are the transformed coordinates
of $a_{1}$ and $a_{2}$ rotated by the angle $\omega$ and $\mu_{a_{1}^{\prime}}$
 and $\mu_{a_{2}^{\prime}}$ are the Gaussian peaks for the distributions in $a_{1}^{\prime}$
and $a_{2}^{\prime}$ respectively and are invariant with the rotation. We can now fit the distributions in both
$a_{1}^{\prime}$ and $a_{2}^{\prime}$ with a Gaussian to characterise
the spread about fitted mean values $\mu_{a_{1}^{\prime}}$ and $\mu_{a_{2}^{\prime}}$,
i.e.
\begin{equation}\label{eqn:rhoa1}
\rho_{1} (a_{1}^{\prime}) = \exp \left \{ -\frac{1}{2} \left (\frac{a_{1}^{\prime}-\mu_{a_{1}^{\prime}}}{\sigma_{a_{1}^{\prime}}}\right)^{2} \right \},
\end{equation}
and
\begin{equation}\label{eqn:rhoa2}
\rho_{2} (a_{2}^{\prime}) = \exp \left \{ -\frac{1}{2} \left (\frac{a_{2}^{\prime}-\mu_{a_{2}^{\prime}}}{\sigma_{a_{2}^{\prime}}}\right)^{2} \right \},
\end{equation}
where $\mu_{a_{1}^{\prime}}$ and $\mu_{a_{1}^{\prime}}$ are as defined previously, and 
$\sigma_{a_{1}^{\prime}}$ and $\sigma_{a_{2}^{\prime}}$ are the characteristic widths of
the Gaussians describing the distributions in $a_{1}^{\prime}$ and
$a_{2}^{\prime}$ respectively. Thus, we obtain,
\begin{equation}
\rho_{a}(a_{1},a_{2}) = \rho_{1}(a_{1}^{\prime})
\rho_{2}(a_{2}^{\prime}) .
\end{equation}

To account for the angular size filtering in 6C* we also require a
functional form which characterises the distribution in projected
linear size $D$. For this we assume that there are no correlations
between $D$ and the other parameters in the small luminosity range we
consider\footnote{We note throughout this section that
our chosen parameterisations are imperfect, e.g. there are
likely to be weak correlations between
radio spectral parameters, linear size and redshift
even over a narrow range of radio luminosity (see
BRW), but attempts to account for such subtleties would
lead to an unacceptable increase in the number of free parameters
while not significantly changing the key results.}. This enables us to model the linear size distribution with a
simple Gaussian distribution in $\log_{10}D$, i.e.
\begin{equation}\label{eqn:rhoD}
\rho_{D} (D) = \exp \left \{ -\frac{1}{2} \left (\frac{\log_{10}D -
\log_{10}D_{\circ}}{\log_{10}D_{1}}\right )^{2} \right \},
\end{equation}
where $\log_{10}D$ is the log of the projected linear size in kpc,
$\log_{10}D_{\circ}$ is the peak of the Gaussian distribution and
$\log_{10}D_{1}$ is the characteristic width of the Gaussian. 

The final factor needed to model the distribution of sources is a
component describing the redshift distribution of the most
luminous radio sources. We use three independent models to achieve
this. 

Model A is parameterised as a single Gaussian distribution in redshift
which is consistent with previous forms describing the evolution in the
comoving space density from the literature (e.g. SH96),
although as noted by JR00 this form has the disadvantage of being a
symmetrical function, thus coupling the low-redshift evolution with
the high-redshift evolution for which there is not necessarily any physical basis. For
model A we take
\begin{equation}\label{eqn:rhozA}
\rho_{A} (z) = \exp \left \{ -\frac{1}{2} \left
(\frac{z-z_{\circ}}{z_{1}}\right)^{2} \right \},
\end{equation}
where $z_{\circ}$ is the redshift of the Gaussian peak and $z_{1}$ is
the characteristic width of the Gaussian.
 
Model B is a 1-tailed Gaussian which becomes constant beyond the
Gaussian peak, this corresponds to a constant comoving space density above
the Gaussian peak redshift, i.e.
\begin{equation}\label{eqn:rhozB}
\rho_{B}(z) = \left\{ \begin{array} {l@{\quad:\quad}l} \exp \left
\{-\frac{1}{2} \left( \frac{z-z_{\circ}}{z_{1}} \right)^{2} \right \}
& z \leq z_{\circ} \\
1.0 & z > z_{\circ} , \end{array} \right.
\end{equation}
where $z_{\circ}$ and $z_{1}$ are as defined previously. 

The third, and final model (C) uses a 1-tailed Gaussian to
parameterise  the low-redshift comoving space
density and a power-law distribution at high redshift, i.e.
\begin{equation}\label{eqn:rhozC}
\rho_{C}(z) = \left\{ \begin{array} {l@{\quad:\quad}l} \exp \left
\{-\frac{1}{2} \left( \frac{z-z_{\circ}}{z_{1}} \right)^{2} \right \}
& z \leq z_{\circ} \\ \left (\frac{1
+ z}{1 + z_{\circ}}\right )^{\eta} &  z > z_{\circ} ,  \end{array}
\right.
\end{equation}
where $z_{\circ}$ is the `break' redshift where the model switches
from the low- to the high-redshift form, $z_{1}$ is the characteristic
width of the half-Gaussian and $\eta$ is the power-law exponent
describing the high-redshift comoving space density. This model has
the advantage of being free to carry on increasing at high redshift,
as we assume there is no {\sl a priori} reason for a decline over the
redshift range of interest. However, it has the disadvantage of an
additional free parameter to the other models. Note that model B is
just model C with $\eta$ fixed at zero.

The distribution for the complete radio luminosity function $\rho$,
is thus given by the product of all these factors, i.e.
\begin{eqnarray}\label{eqn:rlf}
\rho (L_{151},z,a_{1},a_{2},D) = \rho_{\circ} \times \rho_{L}
(L_{151}) \times \rho_{X} (z) \nonumber \\ \times \rho_{a}
(a_{1},a_{2}) \times \rho_{D} (D),
\end{eqnarray}
where $\rho_{\circ}$ is the normalising factor and a free parameter
measured in units of Mpc$^{-3}$ and $\rho_{L}(L_{151})$, $\rho_{X}
(z)$ (for model {\it X}, where {\it X} $\in$ [{\it A}, {\it B}, {\it C}\,] ), $\rho_{a}
(a_{1},\,a_{2})$ and $\rho_{D} (D)$ are dimensionless distribution
functions per $(\Delta \log_{10}L_{151})$, per $(\Delta z)$, per
$(\Delta a_{1}\Delta a_{2})$, and per $(\Delta \log_{10} D)$
respectively.

To find the best-fit values of the free parameters for the various
models, the maximum likelihood method of Marshall et al. (1983) was
used. If $S$ is defined as $-2{\rm ln}\mathcal{L}$, where $\mathcal{L}$ is
the likelihood function, then the aim is to minimise the value of $S$,
which is given by:
\begin{eqnarray}\label{eqn:S}
& S & =-2\sum^{N}_{i=1}\ln [\rho(L_{{151}_{i}},\,z_{i},\,a_{1_{i}},\,a_{2_{i}},\,D_{i})] 
\nonumber \\ & & + 2\int\!\!\int\!\!\int\!\!\int\!\!\int
\rho(L_{151},\,z,\,a_{1},\,a_{2},\,D) \nonumber \\
& & \times  \Omega(L_{151},\,z,\,a_{1},\,a_{2},\,D)
\nonumber \\ & & \times \frac{{\rm d}V}{{\rm d}z} \,{\rm
d}(\log_{10}L_{151})\,{\rm d}z\,{\rm d}a_{1}\, {\rm d}a_{2}\, {\rm d}(\log_{10}D),
\end{eqnarray}
where $\rho(L_{151},\,z,\,a_{1},\,a_{2},\,D)$ is the model
distribution being tested, $\Omega(L_{151},\,z,\,a_{1},\,a_{2},\,D)$
is the sky area available for the samples under consideration in this
paper for a given flux-density, and $({\rm d}V/{\rm d}z)$ is the
differential comoving volume element. The first term is simply the
sum over $N$ sources in the defined sample. The second term is the
integral of the model distribution being tested and should give
$\approx 2N$ for good fits. 

The lower and upper limits of the integral are  $27.63 \leq \log_{10}L_{151} \leq 28.63$ for
cosmology I, and $27.90 \leq \log_{10}L_{151} \leq 28.95$ for
cosmology II, $0 < z < 10$, $-2.2 \leq a_{1} \leq 1.0$, $-0.4 \leq
a_{2} \leq 0.2$ and $-0.3 \leq \log_{10} D \leq 4.0$. To illustrate the
goodness-of-fit of the models the 2-dimensional Kolmogorov-Smirnov
(KS) test, originally proposed by Peacock (1983) was used in the form
of Press et al. (1992). The KS-test enables one to determine the
probability, $P_{\rm KS}$, that a model distribution is a true
representation of the dataset. However, the 2D KS-test may only be
used to reject models, it cannot determine how well a model
fits. Indeed, when the probability, $P_{\rm KS} > 0.2$, its value may
not be accurate but the implication that the model and data are
similar is still correct. For this reason, the relative probability of
each model, $P_{\rm R}$, is also evaluated directly from the
minimisation algorithm.

\subsection{Results of the parametric modelling}\label{sec:resultspara}

\begin{table*}
\begin{center}
\begin{tabular}{c|r|r|r|r|r|r}
\hline\hline 
\mc{1}{c|}{} & \mc{3}{c|}{Cosmology I}  & \mc{3}{c|}{Cosmology II} \\
\mc{1}{c|}{Model} & \mc{1}{c|}{A} & \mc{1}{c|}{B} &
\mc{1}{c|}{C} &  \mc{1}{c|}{A} & \mc{1}{c|}{B} & \mc{1}{c|}{C} \\
\hline\hline
$N$ & 11 & 11 & 12 & 11 & 11 & 12 \\
$\log_{10} \rho_{\circ}$ & $-6.98 \pm 0.12$ & $-7.08 \pm 0.13$ &
$-7.10 \pm 0.12$ & $-7.26 \pm 0.10$ & $-7.33 \pm 0.11$ &  $-7.31 \pm 0.14$ \\
$\beta$ & $2.00 \pm 0.18$ & $2.07 \pm 0.10$ & $1.98 \pm 0.18$ & $2.21
\pm 0.17$ & $2.21 \pm 0.24$ & $2.20 \pm 0.20$ \\   
$z_{\circ}$ & $2.76 \pm 0.22$ & $2.10 \pm 0.23$ & $2.05 \pm 0.30$ &
$2.76 \pm 0.24$ & $2.12 \pm 0.24$ & $2.15 \pm 0.30$ \\
$z_{1}$ & $1.00 \pm 0.13$ & $0.74 \pm 0.16$ & $0.70 \pm 0.17$ & $0.94
\pm 0.13$ & $0.65 \pm 0.14$ & $0.67 \pm 0.15$  \\
$\eta$ & ----- & ----- & $0.10 \pm 1.33$  & ----- & ----- & $-0.06 \pm 1.53$ \\
$\omega$ & $-0.16 \pm 0.01$ & $-0.16 \pm 0.01$ & $-0.16 \pm 0.01$ &
$-0.16 \pm 0.01$ & $-0.16 \pm 0.01$ &  $-0.16 \pm 0.01$  \\
$\mu_{a_{1}^{\prime}}$ & $-0.56 \pm 0.07$ & $-0.55 \pm 0.07$ & $-0.54
\pm 0.07$ & $-0.58 \pm 0.07$ & $-0.59 \pm 0.07$ &  $-0.59 \pm 0.08$ \\
$\sigma_{a_{1}^{\prime}}$ & $0.50 \pm 0.05$ & $0.49 \pm 0.05$ & $0.49
\pm 0.05$ & $0.49 \pm 0.04$ & $0.50 \pm 0.05$ & $0.49 \pm 0.04$  \\
$\mu_{a_{2}^{\prime}}$ & $-0.07 \pm 0.01$ & $-0.07 \pm 0.01$ & $-0.07
\pm 0.01$ & $-0.07 \pm 0.01$ & $-0.07 \pm 0.01$ & $-0.07 \pm 0.01$ \\
$\sigma_{a_{2}^{\prime}}$ & $0.027 \pm 0.003$ & $0.027 \pm 0.002$ &
$0.027 \pm 0.002$ & $0.026 \pm 0.002$ & $0.026 \pm 0.002$ & $0.026 \pm
0.002$ \\
$D_{\circ}$ & $2.04 \pm 0.07$ & $2.02 \pm 0.07$ & $2.00 \pm 0.08$ &
$2.12 \pm 0.07$ & $2.11 \pm 0.07$ & $2.12 \pm 0.07$ \\
$D_{1}$ & $0.59 \pm 0.06$ & $0.58 \pm 0.05$ & $0.59 \pm 0.06$ & $0.56
\pm 0.05$ & $0.56 \pm 0.05$ &  $0.56 \pm 0.05$  \\
\hline
$S_{\rm min}$ & 2803.1 & 2799.8 & 2798.5 & 2941.7 & 2936.7 &
 2936.6 \\
$\ln\sqrt{ {\rm Det} (\nabla\nabla S)}$ & 40.8 & 39.3 &
40.5 & 41.1 & 38.9 & 40.1 \\
$P_{\rm R}$ & 0.04 & 1.00 & 1.11 & 0.01 & 1.00 & 0.32 \\
\hline\hline
\end{tabular}
{\caption{\label{tab:results} Best-fit parameters for the model RLFs
described in Sec.~\ref{sec:paramodels} along with their associated
$1\sigma$ errors. $N$ is the number of free
parameters for each model, $S_{\rm min}$ is the minimum value of
eqn.~\ref{eqn:S}, Det$(\nabla\nabla S)$ is the determinant of the
Hessian matrix evaluated at $S = S_{\rm min}$ and $P_{\rm R}$ is the
relative probability of the models with respect to model B.  A
detailed description of how the errors and relative probabilities are
ascribed can be found in JR00.}}
\end{center}
\end{table*}

\begin{figure}[!ht] 
{\hbox to 0.45\textwidth{ \null\null \epsfxsize=0.45\textwidth
\epsfbox{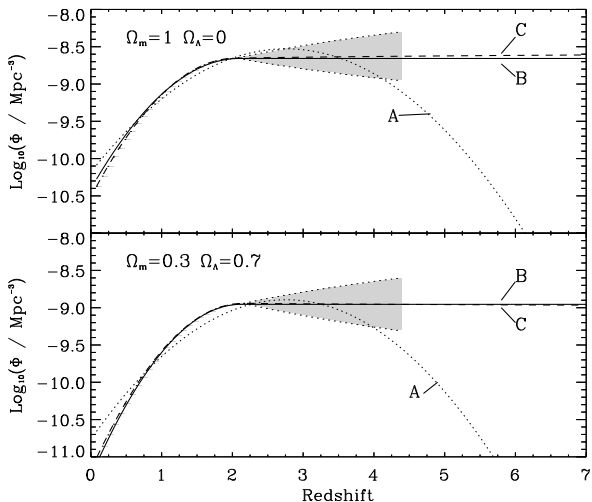}}}
%\epsfxsize=0.45\textwidth
%\epsfbox{rlfscos4ss.ps}}}
{\caption{\label{fig:rlfs} The comoving space density $\Phi$
[$\int\int\int\int\int\rho\,{\rm d}(\log_{10}L_{151}){\rm d}z{\rm
d}a_{1}{\rm d}a_{2}{\rm d}(\log_{10}D)$\,] of the most luminous
low-frequency-selected radio sources using the three models described
in Sec.~\ref{sec:paramodels} for both cosmology I (top) and cosmology
II (bottom). The dotted line is from the best-fit parameters of model
A; the solid line for model B; and the dashed line represents model
C. The shaded region corresponds to the $1\sigma$ uncertainty for the
power-law exponent of model C. This confidence region is uncertain
beyond the limit of our dataset and thus we have excluded it at $z >
4.41$, i.e. beyond the highest redshift source in our samples,
although as the redshift distribution in Sec.~\ref{sec:zdist} shows,
these uncertainties have very little influence on the high-redshift
evolution due to the lack of available comoving volume.}}
\end{figure}

Table~\ref{tab:results} shows the best-fit parameters for the various
models: $S_{\rm min}$ is the minimum of eqn.~\ref{eqn:S} and $P_{\rm
R}$ is the relative probability with respect to model B. The parameter
describing the slope of the luminosity function $\beta$ is consistent with the
previous results of DP90 and Willott et al. (2001) for the
steep-spectrum RLF and Willott et al. (1998) for the RLF of steep-spectrum
quasars. A 2D KS-test to the predicted $L_{151} -z$ plane from the
best-fit parameters give $0.2 \leq P_{\rm KS} \leq 0.4$ for all of our
models in both cosmologies suggesting good working models. The
best-fit values for the correlation between $a_{1}$ and $a_{2}$ are
consistent over all models in the two cosmologies and are also
statistically reasonable with a 2-D KS-test giving $P_{\rm
KS}=0.31$ (Fig.~\ref{fig:a1vsa2}). The parameters describing the Gaussian distribution in $\log_{10}D$ also
span a narrow range between models and KS-probabilities of $P_{\rm KS}
\approx 0.56$ (cosmology I) and $P_{\rm KS} \approx 0.34$ (cosmology II)
suggests adequate fits to the data for all models. The histogram of
$\log D$ / kpc and the fitted Gaussian are shown Fig.~\ref{fig:Dhist}.

\begin{figure}[!ht] 
{\hbox to 0.45\textwidth{ \null\null \epsfxsize=0.45\textwidth
\epsfbox{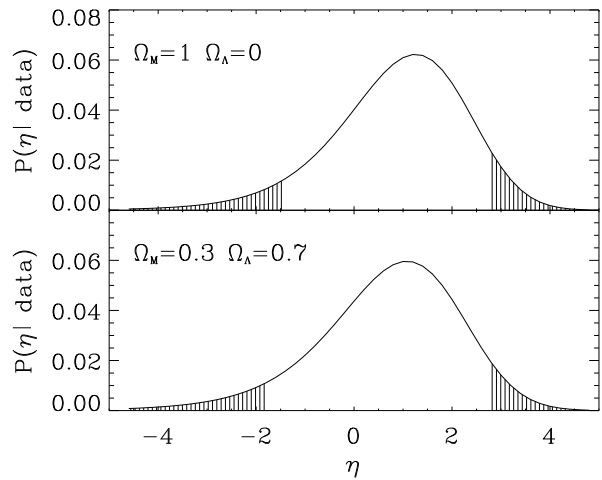}}}
{\caption{\label{fig:etapdf} The probability density function for the
power-law exponent $\eta$ of model C for both cosmology I (top)
and cosmology II (bottom), normalised such that the area under the
curve is unity. The pdf was determined by marginalising over the free parameters
$\rho_{\circ}$, $\beta$, $z_{\circ}$ and $z_{1}$, i.e. the four
parameters with the largest uncertainties from the modelling (all
other parameters were fixed at their best-fit values). The
unshaded region corresponds to the 90\% confidence region.}}
\end{figure}

The parametric models show that model B, with a constant comoving
space density above a critical redshift is $\sim 20$ ($\sim 100$) times
more likely than model A, the model with a symmetric Gaussian decline in
cosmology I (cosmology II). However, one may associate these relative
probabilities with the forced symmetry in model A, thus forcing a poorer
fit at either low- or at high-redshift and these models
must be considered with this property in mind. The best-fit parameters
of model C are probably more relevant in discriminating between models
with sharp cut-offs and those with constant or increasing space
densities.  

Model C uses a 1-tailed Gaussian up to a critical redshift and a
power-law exponent beyond this critical redshift, thus decoupling the
low-redshift from the high-redshift space density. However, this model
requires one more free parameter than models A and B. This introduces
an additional factor when calculating the relative likelihood of this
model, which is described in detail in JR00.
\footnote{On the assumption that most of the prior ranges cancel on dividing the
probabilities of model A and model B, this additional factor is given by, 
\begin{equation}\label{eqn:prior}
\mathcal{F} = (4\pi)^{\frac{11-N}{2}} \Delta \eta ,
\end{equation}
where $N$ is the number of free parameters and the 11 arises from the
number of free parameters in the comparison model (B), and $\Delta
\eta$ is the prior range of the additional parameter. The factor
$\mathcal{F}$ is of order unity and has been neglected for our
analysis here.} The best-fit model suggests that there is very little
evidence for an abrupt decline in the comoving space density of the
most luminous low-frequency-selected sources at high redshift. Indeed
the best-fit to the power-law exponent in cosmology I
(Table~\ref{tab:results}) suggests a steady increase up to an
indeterminable redshift, and the best-fit in cosmology II is a
power-law exponent of $\sim 0$ thus giving a constant comoving space
density, albeit with an uncertainty encompassing moderate declines and
inclines. This uncertainty on the power-law exponent $\eta$ is
represented by the shaded region in Fig.~\ref{fig:rlfs} and also by a
probability distribution function (pdf) in Fig.~\ref{fig:etapdf} and
it is obvious that there exists a large range in which $\eta$ may
reside. This behaviour is different from the pdf for the power-law
exponent from the parametric modelling of the flat-spectrum population
of JR00 at $\approx 2-3\sigma$ level, where the pdf derived from the
$V/V_{\rm max}$ method peaked at $\eta \approx -3$, which is also
consistent with their best-fit parametric models.

The dissimilarity between the behaviour observed by
JR00 for the most luminous flat-spectrum sources from a high-frequency
selected sample, where there was $\sim 2\sigma$ evidence for a decline
by a factor of $\sim4$ between $2.5 \leq z \leq 5$, and this analysis is
resolvable within the large uncertainties. Although the steep decline
ruled out at the $2\sigma$ level for flat-spectrum sources in JR00 is
ruled out at roughly the 4$\sigma$ level here. However, if the
difference in the best-fit power-law exponents is a real difference
between the two radio source populations then it may not necessarily
be due to a breakdown of orientation-based unified schemes and we will
discuss this in Sec.~\ref{sec:discussion}.

\subsection{The effect of changing the lower luminosity limit}\label{sec:changelum}

We now investigate the effect that the positioning of the lower limit
in radio luminosity has on the outcome of the parametric
modelling. Each model was re-fitted with both the lower-limit in radio
luminosity increased by 0.25 dex and also lowered by 0.25 dex. The
relative probabilities of the best-fit models in both cosmologies
changed very little with this alteration. The uncertainties on the
fitted parameters increased when the luminosity limit was raised, as
one might expect due to the small number statistics becoming
increasingly important. The steepness of the RLF, governed by the
$\beta$ parameter in our models was consistently $\approx 2.1$ within
the uncertainties, suggesting that the steepness of the RLF in the
high-luminosity regime is consistent with $\beta$ at lower
luminosities. The power-law exponent $\eta$ for the high-redshift
evolution in model C also varied very little from a best-fit value of
$\approx 0$, although the large errors associated with this parameter
still make shallow declines and also shallow increases easily
plausible. Therefore, we conclude that the positioning of the
lower-limit in radio luminosity has an insignificant effect on the
relative probabilities and the fitted values of the parametric
modelling if the modelling is restricted to only the most radio
luminous sources, where we are confident of completeness in all of the
samples.

To test the susceptibility of our models to small number statistics we
also re-fitted our models with an additional {\it fake} $z = 6$
source. Inclusion of the $z = 6$ radio source in our models also
has very little effect on the relative probabilities, although model A
becomes slightly less plausible relative to model B in both
cosmologies as one would expect. The $\eta$ parameter is still consistent with a value
of $\eta \approx 0$ and no positive evolution in the comoving space
density is required (although the best-fit values are $\eta = 0.35$
and $\eta = 0.20$ for cosmology I and II respectively). The peak of
the low-redshift Gaussian is constrained at a value of $z_{\circ} \sim
2$, although the errors on this
parameter are high and correlated with the $\eta$ parameter. We therefore
conclude that including a radio galaxy at $z = 6$ has little effect
on the overall conclusions of our main analysis, as the small number
statistics involved in including a single source at high redshift are
easily compatible with the original best-fit models without the $z=6$
source.  This is also highlighted by the predicted redshift
distribution from our original model C (Fig.~\ref{fig:zcumdist_pdf}), where the presence of a single
source at $z \sim 6$ could easily conform to the predictions in both
cosmologies.

\section{The banded $V / V_{\rm max}$ statistic}\label{sec:vvmax}

\begin{figure}[!ht] 
{\hbox to 0.45\textwidth{\epsfxsize=0.45\textwidth
\epsfbox{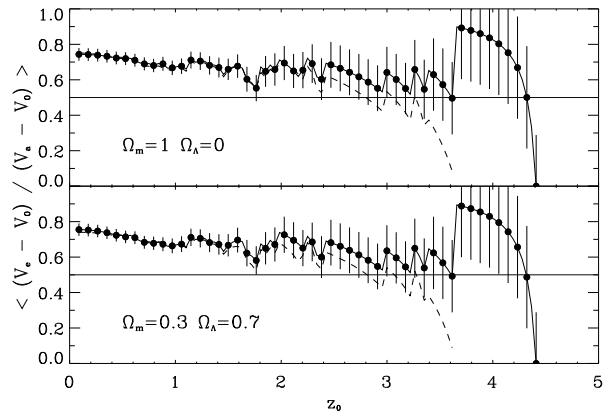} }} 
{\caption{\label{fig:vvmax} The banded $V/V_{\rm max}$ statistic for
the most radio luminous steep-spectrum sources considered in this
paper. The solid line corresponds to the statistic with the lensed
radio galaxy 6C*0140+326 included in the analysis and the dashed line
is the statistic excluding this source. The solid vertical lines
correspond to the $1\sigma$ errors on the statistic assuming $\sigma =
1/\sqrt{12N}$ (Avni \& Bahcall 1980), where $N$ is the number of
sources above $z_{\circ}$. }}
\end{figure}

It is also useful to view our parametric modelling along with a
non-parametric form in which the behaviour in the space density of the
most luminous, high-redshift sources does not depend on the assumed
parameterisation. The most common method for this non-parametric
analysis is the $V/V_{\rm max}$ method in its banded form (Avni \&
Bahcall 1980; see also DP90, and JR00). Unfortunately, with a filtered
sample such as 6C* the actual significance of any value of $V/V_{\rm
max} > 0.5$ may be viewed as inconsequential due to the bias towards
finding objects at high redshift. However, if this banded test shows a
significant decline in the comoving space density of sources with a
sample biased to finding objects at high redshift then it would prove
to be a significant detection of any redshift
cut-off. Fig.~\ref{fig:vvmax} shows the banded $V/V_{\rm max}$
statistic for all of the sources in the top-decade in radio
luminosity considered in our parametric modelling. It is apparent that
the statistic shows no significant evidence for a decline in the
comoving space density out to $z \sim 4.5$, if the highest-redshift
(lensed) radio galaxy 6C*0140+326 is included. If we now repeat the
analysis without this object there also appears to be little evidence
of a decline, with all points lying within $1\sigma$ of the line of
no-evolution at $V/V_{\rm max} = 0.5$. Therefore, we find no evidence
for a decline in the comoving space density of the most luminous
radio sources using this non-parametric approach, in agreement with
the best-fits from our parametric models. The high values of $V/V_{\rm
max}$ in Fig.~\ref{fig:vvmax} reflect a combination of genuine
positive evolution at $z < 2$ as the 3CRR sources dominate in this
regime, and biases due to selection methods at higher redshifts where
the sample becomes increasingly dominated by 6C*.

\section{Predicted redshift distributions}\label{sec:zdist}

\begin{figure}[!ht] 
{\hbox to 0.45\textwidth{\epsfxsize=0.45\textwidth
\epsfbox{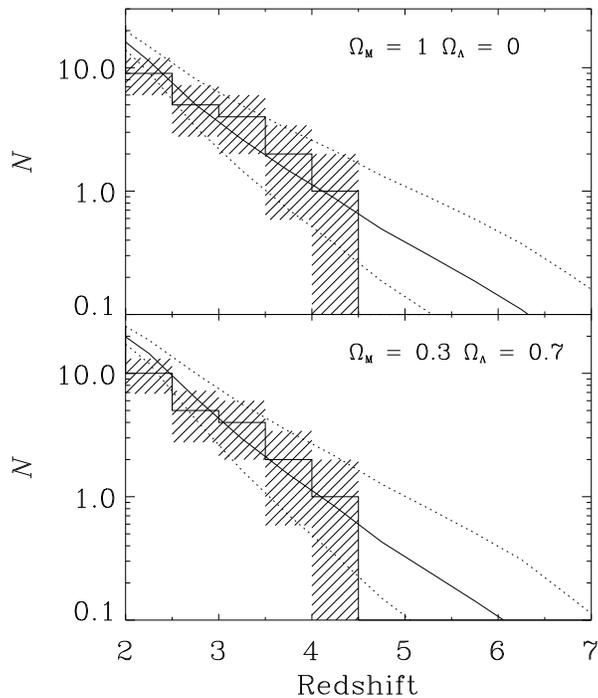} }} 
{\caption{\label{fig:zcumdist_pdf} Cumulative redshift distributions at $z
> 2$ for model C in cosmology I (top) and cosmology II (bottom). The
histogram represents the actual cumulative redshift distribution of
the sources in the 3 samples used, the hashed regions are the
$\sqrt{N}$ errors on the binned data. The solid curve is the
predicted distribution from the best-fit values of model C. The two
dotted curves represent the curves for the $1\sigma$ uncertainty on
the high-redshift exponent $\eta$, using the uncertainties generated in our
parametric modelling.}}
\end{figure}

We now consider the redshift distribution expected given our best-fit
values for model C. Fig.~\ref{fig:zcumdist_pdf} shows the cumulative
redshift distribution for our sample for $z > 2$, along with the
predicted distribution for model C from our best-fit model. The forced
symmetry in the estimation of these errors by fitting a symmetrical
Gaussian to the minima in the parametric modelling (see JR00) means
that the uncertainties may be over-estimated on one side and
underestimated on the other. Therefore, although the predicted
redshift distributions are in accord with the data the actual
$1\sigma$ errors may not be telling us the whole story. Thus, the pdf
shown in Fig.~\ref{fig:etapdf} is probably a better representation of
the uncertainties in the parameter space we are most interested in. We
can see that the model parameters are consistent with the data all the
way through the high redshift regime. The upper limit on the
$\eta$ parameter from our 1$\sigma$ uncertainty is easily consistent
with the data. This highlights the effect of the small number
statistics under consideration at the highest redshifts. The steepness
of the RLF, combined with the sharp reduction in the available
comoving volume due to the flux-density limits of the samples means
that, even without any decline in the comoving space density of radio
sources, the number of sources one would expect to find is $< 1$.

%\section{Discussion}\label{sec:discussion}
\section{The RLF of steep- and flat-spectrum radio sources
in the context of unified schemes}\label{sec:discussion}

Our work on the radio luminosity function of both steep-spectrum radio
sources (this paper) and
flat-spectrum radio sources (JR00) has highlighted the difficulty in
establishing the comoving space density of these sources at redshifts
which push to the limits of flux-density-limited radio
samples. Although the uncertainties on the fitted parameters
describing the high-redshift behaviour of both these types of radio
source are quite large they are markedly different in their mean value
for the high-redshift power-law exponent. Indeed the evidence from our
parametric modelling favours a steady decline for the flat-spectrum
population by a factor $\sim 4$ between $2.5 \leq z \leq 5$ (JR00), 
whereas analysis of the low-frequency samples in this paper lead us to
a best-fit model with a roughly constant comoving space beyond a peak
redshift around $z \approx 2.5$.

This discrepancy is just resolvable within the uncertainties of both
models, so it is plausibly a statistical fluke that we are seeing
different evolutionary behaviour between the two populations of radio
sources. However, other plausible explanations exist and should be
investigated further. 

The most obvious difference between the flat- and steep-spectrum
samples is the selection frequency. The high selection frequency
(2.7\,GHz) for the flat-spectrum sample means that the majority of
objects are core-dominated Doppler-boosted quasars and, as argued by
JR00, Giga-Hertz Peaked spectrum (GPS) sources may also make an
important contribution. If a significant fraction of the most luminous
flat-spectrum sources are indeed GPS sources, then at an observed
frequency of 2.7\,GHz, the radio spectrum of a GPS source may have
steepened by an amount sufficient for it to drop below the sample flux-density
limit or for it to have been rejected from the sample on account of its
steep observed spectral index ($\alpha > 0.5$) if it is at
high redshift. Indeed even if the core-dominated, Doppler-boosted
sources are a dominant part of the top-decade of the flat-spectrum
luminosity function, and assuming a conservative estimate for the
fraction of GPS sources, a decline in the comoving space density
would be observed at high redshift. This is because of the drop off in
GPS sources due to K-correction effects, allowing a roughly constant
comoving population of Doppler-boosted sources. This point was made
by JR00 and they suggest a model for the RLF is needed in which both
the Doppler-boosted and GPS populations are explicitly considered. The banded $V/V_{\rm max}$ method used by JR00 was able to
account for the spectral shape of individual sources and although
small number statistics make it very difficult to be certain (see the
broad 90\% confidence region in their Fig.9b), the presence of some
high-redshift decline in one or more of the Doppler-boosted and GPS
populations appears robust to spectral-dependent selection effects.

Another difference between the flat- and steep-spectrum sources is the
steepness of the luminosity function in the top decade. JR00 found
values in the range $1.2 \le \beta \leq 1.75$ for the top decade of
the flat-spectrum population, whereas values of $\beta \geq 2.0$ have
been determined for the top decade of steep-spectrum RLF in this
analysis and over a broader range in luminosities in previous work
(e.g. Willott et al. 1998; Willott et al. 2001). DP90 also found a
marginally steeper luminosity function for their steep-spectrum population
($\beta \sim 2.2$) compared to their flat-spectrum sample ($\beta \sim
2.0$).

The difference in both the high-redshift decline and values of $\beta$
between the flat- and steep-spectrum sources can be resolved by
considering the concave shape of the RLF (Willott et al. 2001). The
most luminous flat-spectrum population probes a different part of the
underlying RLF than the most luminous low-frequency-selected sources
(JR00). Thus, if the position of high-frequency selected flat-spectrum
sources on the underlying RLF are at luminosities which are lower than
those of the most luminous low-frequency-selected sources and lie on
or near the `break' in the RLF at $\log_{10} L_{151} \sim 26.5$ for
cosmology I (Willott et al. 2001), then this automatically supports
the steeper values derived for $\beta$ from our steep-spectrum
samples. In other words, the slope in the RLF steepens with increasing
intrinsic luminosity, so either Doppler boosted or GPS sources in
high-frequency samples will be drawn from a population where the slope
in the underlying RLF is flatter. Indeed, it seems likely that the
Doppler boosted and GPS objects are drawn from different points on the
underlying RLF, so there may well be differential effects: JR00 for
example, have suggested this as the cause of an increasing fraction of
GPS sources with increasing luminosity in high-frequency selected,
flat-spectrum samples.

The difference in the evolution of high-redshift comoving space
density between the flat- and steep-spectrum populations can now be
understood if there is a decline in the
comoving space density at high redshift for lower-luminosity
low-frequency sources. This would be very important to confirm because
the luminosity density ($\int L\rho$\,${\rm d\log_{10}}L$) of the
population is dominated by the space density near the break of the
RLF. It is impossible to probe this break region directly with
the complete flux-density-limited samples currently available. 
%The
%results from the flat-spectrum population (DP90 \& JR00) are
%tentative, but supply indirect evidence that the luminosity density is
%in decline at high redshift.

We therefore conclude that, although the high-redshift space density
and the steepness of the RLF may be different for flat- and
steep-spectrum sources in the top-decade of luminosity, this can easily
be explained by some combination of two effects: first, a mixed
population of Doppler boosted and GPS sources in the high-frequency selected
flat-spectrum samples; second, and arguably more importantly, the
flat- and steep-spectrum populations lie on different parts of an
underlying RLF which is concave (Willott et al. 2001). Thus,
orientation-based unified schemes are not challenged by the comparison
of constraints on the high-redshift evolution of flat- and
steep-spectrum radio sources.

\section{Conclusions}\label{sec:conclusions}
The main conclusions we have drawn from our parametric
modelling analysis of the most luminous low-frequency-selected radio
sources are:
\begin{itemize}
\item Their low-redshift evolution follows the same track as that of
flat-spectrum radio sources, optically selected AGN (e.g. Schmidt,
Schneider \& Gunn 1995) and X-ray selected AGN (e.g. Miyaji, Hasinger
\& Schmidt 2000), with an increase in comoving space density by a
factor $\gtsim 100$ over $0 \leq z \leq 2.5$.
\item Their steep increase in space density slows down at $z
> 2.5$ in spatially-flat cosmologies with either $\Omega_{\Lambda} = 0$ or
$\Omega_{\Lambda} =0.7$.
\item The saturated Gaussian model B with a constant comoving space
density beyond a peak redshift, is $\approx 25$ ($\approx 100$) times
more likely than model A with a Gaussian rise and cut-off for cosmology I (II).
Although this ratio may partly reflect the forcing of the low-redshift increase to be symmetric with the high-redshift decline in
the comoving space density of the most luminous radio sources, it
does illustrate the significant difference between radio-loud AGN and
optically selected quasars [which do have a roughly Gaussian
evolutionary behaviour, e.g. Schmidt et al. (1995); Kennefick,
Djorgovski \& de Carvalho (1995)].
\item The best-fit model is consistent with a constant comoving space
density beyond the peak at $z \approx 2.5$ to an indeterminable
redshift, although the uncertainties span a wide range covering
both moderate declines and continuing shallow inclines.
\item The difference between the high-redshift space density for flat-
and steep-spectrum sources is marginally significant, although there
remain worries concerning the mixture of Doppler Boosted and GPS
sources in the flat-spectrum population. An explanation exists in
which the high-frequency flat-spectrum sources are probing the
underlying RLF at fainter intrinsic luminosities than the
top-decade discussed in this paper. This could also resolve
the difference in $\beta$ between the two populations, because at
lower-luminosities the sources are edging towards the break in the
luminosity function where $\beta$ becomes flatter (Willott et
al. 2001). This is important to confirm since the tentative evidence
for a decline in the flat-spectrum population supplies indirect
evidence of a decline in the luminosity density in the radio source
population at high redshift.
\end{itemize}

\section*{Acknowledgements}
We are grateful to a large number of colleagues who
have devoted effort and observing time to the quest for
redshifts for the 6CE and 6C* samples.
This research has made use of the NASA/IPAC Extra-galactic Database, which
is operated by the Jet Propulsion Laboratory, Caltech, under contract
with the National Aeronautics and Space Administration.
MJJ thanks PPARC for the receipt of a studentship.

\end{document}